\documentclass{skaox2006}
\usepackage{graphicx}
\begin{document}
   \title{Expanding e-MERLIN with the Goonhilly Earth Station}



   \author{I. Heywood$^{1}$, H.-R. Kl\"{o}ckner$^{1,2}$, R. Beswick$^{3,4}$, S. T. Garrington$^{3,4}$, J. Hatchell$^{5}$, M. G. Hoare$^{6}$, 
   M. J. Jarvis$^{7}$, I. Jones$^{8}$, T. W. B. Muxlow$^{3}$
          \and
          S. Rawlings$^{1}$
          }

   \institute{$^{1}$Subdepartment of Astrophysics, University of Oxford, Denys-Wilkinson Building, Keble Road, Oxford, OX1 3RH, UK\\
         	$^{2}$Max-Planck-Institut f\"{u}r Radioastronomie, Auf dem H\"{u}gel 69, 53121 Bonn, Germany\\
		$^{3}$e-MERLIN / VLBI National Radio Astronomy Facility, Jodrell Bank Observatory, The University of Manchester, Macclesfield, Cheshire, SK11 9DL, UK\\
		$^{4}$Jodrell Bank Centre for Astrophysics, School of Physics and Astronomy, The University of Manchester, Oxford Road, Manchester, M13 9PL, UK\\
		$^{5}$Astrophysics Group, CEMPS, University of Exeter, Stocker Road, Exeter, EX4 4QL, UK \\
		$^{6}$School of Physics and Astronomy, University of Leeds, Leeds, LS2 9JT, UK\\
		$^{7}$Centre for Astrophysics Research, STRI, University of Hertfordshire, Hatfield, AL10 9AB, UK\\
		$^{8}$Goonhilly Earth Station Ltd, Goonhilly Downs, Helston, Cornwall, TR12 6LQ, UK\\
             }

   \abstract{A consortium of universities has recently been formed with the goal of using the decommissioned telecommunications
   infrastructure at the Goonhilly Earth Station in Cornwall, UK, for astronomical purposes. One particular goal is the introduction
   of one or more of the $\sim$30-metre parabolic antennas into the existing e-MERLIN radio interferometer. 
   This article introduces this scheme and presents some simulations which quantify the improvements that would be brought
   to the e-MERLIN system. These include an approximate doubling of the spatial resolution of the array, an increase in its N-S extent
   with strong implications for imaging the most well-studied equatorial fields, accessible to ESO
facilities including ALMA. It also increases the overlap
   between the e-MERLIN array and the European VLBI Network. We also discuss briefly some niche science areas in which
   an e-MERLIN array which included a receptor at Goonhilly would be potentially world-leading, in addition
   to enhancing the existing potential of e-MERLIN in its role as a Square Kilometer Array pathfinder instrument.}
   \authorrunning{Heywood et al.}
   \titlerunning{Expanding e-MERLIN with the Goonhilly Earth Station}
   
   \maketitle
%
%

\section{Introduction}

The Goonhilly Earth Station (GES; lat = 50.0504 N, lon = 5.1835 W) is 
home to three large ($\sim$30-metre) parabolic antennas which were 
originally built for telecommunications. GES was in fact the world's 
first satellite communication station, its original parabolic antenna 
being used to receive the first transatlantic television broadcast via 
the Telstar satellite approximately half a century ago.

   \begin{figure}
   \centering
\includegraphics[width=\columnwidth]{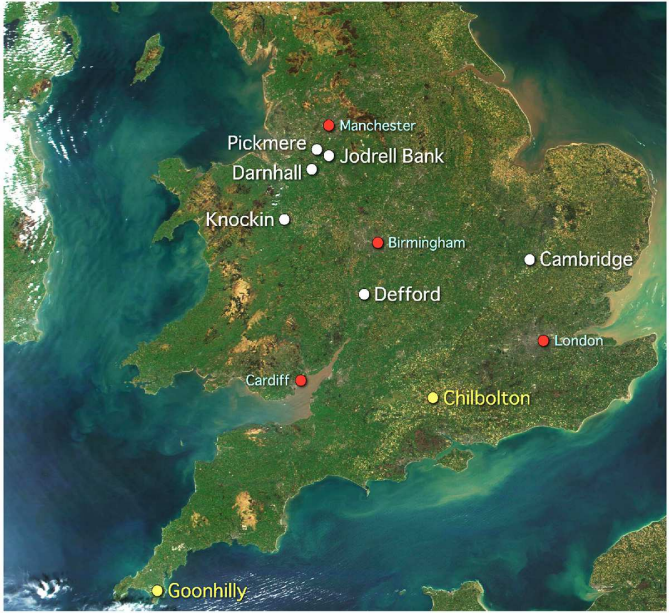}
\caption{The locations of the existing telescopes of the e-MERLIN array and the location
   of the Goonhilly Earth Station. The additional long baselines provided by a station at Goonhilly
   are immediately apparent.}
            \label{fig:merlinmap}
   \end{figure}
   
Since most satellite communication and international telecommunication needs are now met using
a combination of much smaller dishes and undersea cables, 
several of which terminate at the Goonhilly site, the satellite operations 
using the large dishes at Goonhilly were ceased in 2008. The Consortium of Universities for 
Goonhilly Astronomy (CUGA\footnote{{\tt 
http://www.ast.leeds.ac.uk/$\sim$mgh/CUGA.htm}}; consisting of 
the universities of Exeter, Hertfordshire, Leeds, Manchester and Oxford) 
has recently formed in order to use the existing infrastructure at 
Goonhilly for astronomical purposes. One such application is the 
connection of one or more of the large parabolic antennas into the e-MERLIN 
array.

In this short article we present some simulations to demonstrate the 
improved performance of an e-MERLIN which includes an antenna at 
Goonhilly. We also discuss some scientific applications which would be 
ideal for such an instrument. This article is also a follow-up to a previous paper which discusses the benefits 
of the inclusion of the 25-metre parabolic antenna at Chilbolton Observatory (Heywood et al., 2008).

The e-MERLIN array, which is described in more detail in Section 2, serves as the UK's Square Kilometre Array (SKA) 
pathfinder facility. It has a unique science scope due to its 
``intermediate'' baseline lengths (i.e. occupying the gap between those of 
the EVLA and VLBI networks such as the VLBA and EVN). Its scientific 
exploitation can thus potentially strongly influence the eventual 
placement of the dishes in Phase-I of the SKA (SKA$_{1}$; Garrett et 
al., 2010). Goonhilly may be particularly important in this respect as 
it enhances the resolution of e-MERLIN enough to allow the exploration of 
specific SKA Key Science aspects which are certain to require such 
intermediate baseline lengths, but for which there is little direct 
observational evidence concerning just what range of baselines is 
needed.

The high cost of data transport from outlying dishes (i.e. the receptors 
which form the intermediate and longest baselines in the SKA) means that 
e-MERLIN will also inform on the trade-off between science return and 
project cost for SKA$_{1}$.

\section{e-MERLIN}

The e-MERLIN\footnote{{\tt http://www.e-merlin.ac.uk}} array, operated 
by the University of Manchester at Jodrell Bank Observatory, is a radio 
interferometer consisting of seven radio telescopes situated around the 
UK, with a current maximum baseline of approximately 217 km. Upgrades to 
the original MERLIN array include new receivers and a 
high-bandwidth fibre link to transport the data back to a new digital correlator. 

The upgraded system allows e-MERLIN to deliver micro-Jansky 
sensitivity with up to 4 GHz of instantaneous bandwidth at L- (1.3 - 1.8 
GHz), C- (4 - 8 GHz) and K-band (22 - 24 GHz), with an angular 
resolution of 10 - 150 milliarcseconds.

The locations of the existing e-MERLIN telescopes are shown by the white 
markers on Figure \ref{fig:merlinmap}, and also marked on this map in 
yellow are the locations of the GES and Chilbolton Observatory. Note 
that Jodrell Bank hosts two antennas: the 76-metre Lovell Telescope and 
the 25-metre Mark-II. The extra long baselines formed by connecting a 
receptor at Goonhilly to each of the existing e-MERLIN telescopes should 
be immediately apparent.

The surface accuracy of the Lovell telescope precludes its inclusion in K-band
observations, and similarly the telescope at Defford cannot be used because of the
spacings of its mesh reflector. 

Note that even if both the Lovell and Mark-II telescopes are included in an e-MERLIN
observation there is still one spare correlator input which could be fed by
an additional antenna.

\section{Extra baselines and improved $uv$-coverage}

The GES has three large antennas on site: Goonhilly-1 (or ``Arthur''; 
25.9 m diameter), Goonhilly-3 (or ``Guinevere''; 29.6 m diameter) and 
Goonhilly-6 (or ``Merlin'', no relation; 32 m diameter). We assume that the proposed 
extension of e-MERLIN would initially involve the use of Goonhilly-1 at 
L- and C-bands, and the eventual use of Goonhilly-3 at C- and K-bands.

In an interferometric observation each baseline as projected on the sky 
measures a single Fourier component of the sky brightness distribution 
at a particular observing frequency. A single correlation product 
between two antennas is the atomic unit of an interferometric 
observation, a ``visibility", and this complex number and its 
conjugate occupy distinct points in the $uv$-plane. The longest baseline 
of an array occupies the outermost region of the $uv$-plane and provides 
the highest resolution. Similarly the shortest baseline samples the 
largest spatial scales and occupies the inner-most region. The shortest 
baseline also determines the spatial scale at which objects begin to get 
``resolved out''.

In addition to increasing the sensitivity via an increase in collecting 
area, adding more antennas to an array increases the range of spatial 
scales it can sample by virtue of the the extra baselines which are 
created, thus adding more points to the $uv$-plane coverage. 
Earth-rotation increases the $uv$-coverage further by changing the 
projection of the baselines on the sky with time, causing the baselines 
to trace arcs on the $uv$ plot. Multi-frequency synthesis can also be 
used to increase the spread of measurements in the $uv$-plane, by 
observing over a broad bandwidth, which manifests itself as a radial 
spread in the $uv$ plot.

We can demonstrate the improved $uv$-plane coverage that the 
introduction of Goonhilly would bring by simulating an e-MERLIN 
observation with an additional antenna at the location of Goonhilly-1. 
This is accomplished by generating a set of simulated 
visibilities.\footnote{Simulations are performed using the {\tt sm} tool 
within the {\tt CASA} ({\tt http://casa.nrao.edu}) package.} The antenna 
table is constructed using the locations of antennas from an actual 
e-MERLIN observation with an additional entry for the Goonhilly-1 
antenna.


\begin{figure*}
\vspace{400pt}
\includegraphics{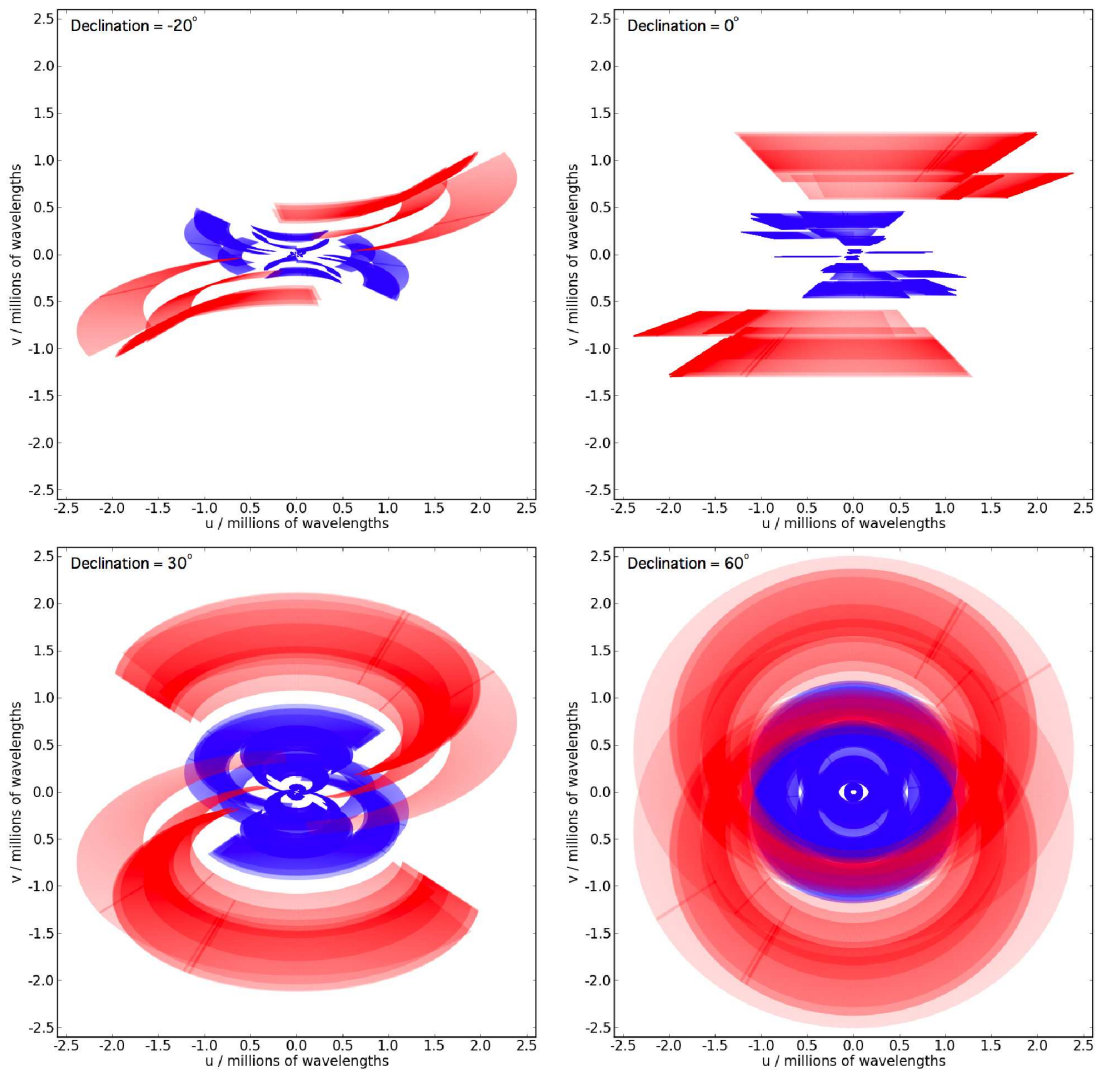}
\caption{Four full synthesis $uv$-coverage plots for the e-MERLIN array with
      a frequency range of 1.2 to 1.712 GHz, for four source declinations as indicated on
      each panel. The extra baselines provided by the Goonhilly station are shown in red.
      \label{fig:uvplots}}

\end{figure*}

We simulate a full earth-rotation L-band observation with 256 channels 
between 1.2 and 1.712 GHz for three representative declinations, and an 
antenna elevation cut-off of 8 degrees is also imposed.

Figure \ref{fig:uvplots} shows the $uv$-plane coverage of the e-MERLIN 
array with the inclusion of the Goonhilly antenna. This figure is 
constructed by plotting the resulting $u$ and $v$ coordinates from the 
simulated visibilities at four representative declinations and colouring in red any baseline that is formed 
with the Goonhilly-1 antenna.

These plots clearly demonstrate the increased and improved spatial frequency sampling and resolution 
advantages of an extended e-MERLIN array. Of particular note is the upper panel showing the declination = 0 
case. When observing the celestial equator with an interferometer the 
baseline tracks are in a purely E-W direction, generally resulting in 
undesirable N-S structure in the dirty beam (or point-spread function). 
Any N-S extension of the array itself goes some way to alleviating this 
problem. 

Note that for the simulations presented here we include the Mark-II but 
not the Lovell Telescope. Generally only one of these two telescopes is 
included in the interferometer, and inclusion of the Lovell instead of 
the Mark-II would make little difference to the $uv$-plane coverage as 
the telescopes are located very close to one another. The main effect 
the Lovell Telescope has when included in e-MERLIN (at L- and C-band) is 
to form a series of very sensitive baselines due to its 76-metre 
diameter, increasing the sensitivity of the array by a factor of 2--3.

Although the inclusion of the Lovell has implications for wide-field 
observing due to its comparatively small field-of-view, the proposed 
e-MERLIN Legacy Surveys will generally benefit from its inclusion, and 
such programmes especially those
targeting faint objects at high resolution, would be enhanced greatly by the very sensitive 
long-baseline formed by Goonhilly and the Lovell (Kl\"{o}ckner et al., 2011, in prep.)

\section{Improved beam shapes}
\label{sec:beamsection}

The visibilities used to generate the $uv$ plots in the previous section are imaged with natural
weighting to determine the dirty beam (i.e. point-spread function) of a given observation\footnote{When imaging an 
interferometric observation the visibilities are gridded and then 
Fourier transformed. The $uv$-plane coverage and the dirty beam are 
directly coupled, in that they are Fourier transforms of one-another. Natural weighting means the gridded
visibilities are weighted according to the density of points in a $uv$ cell. Since there are always more short
baselines than long ones naturally-weighted visibilities result in a \emph{lower} resolution than that returned
by different weighting functions. Uniform weighting applies equal weight to all visibilities and generally results
in the highest resolution at the expense of imaging sensitivity.} 
The generation of these beams naturally incorporates multi-frequency synthesis.

A two-dimensional Gaussian is fitted to the central lobe of each of the generated beam patterns
and the results of this procedure are presented in Table \ref{tab:resolutions}. In addition to the major
and minor axes of the Gaussian their ratio is also tabulated. This provides a measure of the ``circularity''
of the beam.

Having a point-spread function which is more circular for equatorial
fields will benefit survey
science in general, as many of the best-studied multiwavelength fields are situated on or near the
celestial equator, and it is likely to be the site of the deepest future ALMA observations. 
Circular beams are also of particular benefit to weak lensing
experiments.
In order to overcome the systematic limits in deep, wide-area weak
lensing surveys, and make the first steps towards exploiting weak lensing
with the SKA (Blake et al., 2004), there is an urgent
need to extend existing first attempts at radio-optical weak lensing surveys
(Patel et al., 2010) to cover larger-area weak-lensing surveys, typically situated
at or below the celestial equator. 

\begin{table}
\centering
\caption{Major axis, minor axis and position angle of the Gaussians fitted to the central lobes of the naturally-weighted  dirty beams for full-track observations
at four declinations at L-band. 
These effectively correspond to the resolution of the array. Also included in the table are the ratios of the major to minor axis
for each scenario. The closer this number is to unity the more circular the beam, and Goonhilly provides
vastly more circular beams for low |Dec| observations. (eM = the existing e-MERLIN array; eM+G includes Goonhilly-1). 
\label{tab:resolutions}}
\begin{tabular}{lllll}
\hline
Declination          		& B$_{maj}$ & B$_{min}$ & PA     & B$_{maj}$/ \\
(and array)			& (arcsec)  & (arcsec)  & (deg)  & B$_{min}$  \\ \hline
$\delta$ = -20$^{\circ}$ (eM)	& 0.688     & 0.159     & 12.72  & 4.33       \\
$\delta$ = -20$^{\circ}$ (eM+G) & 0.352     & 0.118     & 163.482& 2.98       \\
				& 	    &           &        & \\
$\delta$ = 0$^{\circ}$ (eM)	& 0.393     & 0.166     & 22.28  & 2.37\\
$\delta$ = 0$^{\circ}$ (eM+G)	& 0.195     & 0.116     & -29.65 & 1.68\\
				& 	    &           &        & \\
$\delta$ = +30$^{\circ}$ (eM)	& 0.225     & 0.186     & 23.45  & 1.21\\
$\delta$ = +30$^{\circ}$ (eM+G)	& 0.135     & 0.120     & 134.39 & 1.13\\
				& 	    &           &        & \\
$\delta$ = +60$^{\circ}$ (eM)	& 0.200     & 0.183     & 0.74   & 1.09\\
$\delta$ = +60$^{\circ}$ (eM+G)	& 0.133     & 0.126     & -1.06  & 1.06\\
\hline
\end{tabular}
\end{table}

As mentioned previously, the 
baselines during a declination = 0 observation move through the $uv$-plane in a purely E-W direction as can 
be seen in Figure \ref{fig:uvplots}. This generally results in unfavourable 
N-S sidelobes in the dirty beam, and increasing the N-S extent of e-MERLIN by including Goonhilly mitigates this problem 
significantly.

\section{Applications}

Here we briefly discuss some e-MERLIN science cases for which the 
enhanced performance of the instrument would be particularly valuable.

\subsection{Strong gravitational lenses (primarily L-band)}

The concomitant advent of an efficient means of finding 
strongly-gravitationally-lensed systems (HERSCHEL; \cite{negrello}) and a 
means of following these up at high spatial resolution (e-MERLIN; Deane 
et al.~2011, in prep.) opens up some fascinating new possibilities. 
Goonhilly provides e-MERLIN with ``HST-like'' resolution even at L-band 
where the lensed radio sources are typically brightest. As emphasised by 
\cite{dalal} and \cite{vegetti}, such resolution is sufficient and 
necessary to adequately probe the dark matter distribution of 
quad-system gravitational lenses down to sub-galactic scales. 
Critically, the new HERSCHEL / e-MERLIN method will target lensed 
sources that are intrinsically extended enough to avoid microlensing 
effects, but compact enough to be noticeably perturbed by small-scale 
substructure in the dark matter. By measuring the mass fraction in 
sub-structure in a large sample of new gravitational lenses, e-MERLIN 
with Goonhilly could prove to be the current-generation facility 
best-suited to testing the predictions of the CDM model on mass scales 
well below that of the Milky Way satellites.

\begin{figure*}
\vspace{360pt}
\includegraphics{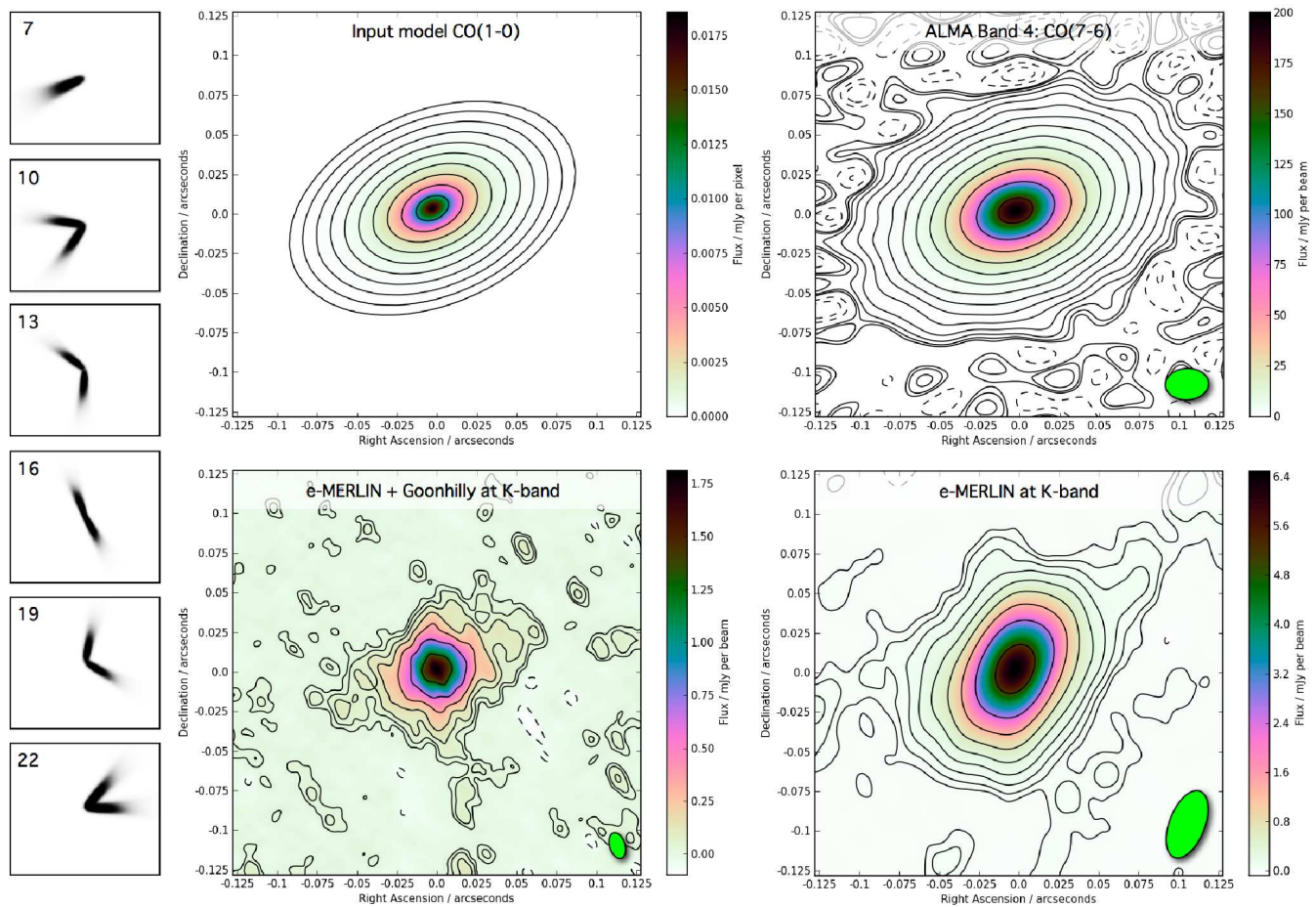}
\caption{Simulations of observations of a CO disk at a redshift of 4.023.
      Simply to illustrate the channelised source morphology, the left-hand column shows thumbnail images of a selection of numbered channels from the input model, which is
      the molecular gas disk of a simulated galaxy at z~=~4.023 as described by the S$^{3}$-SAX simulation.
      The four, large colour frames show the frequency-averaged cube emission for the input model (upper-left),
      an ALMA Band 4 observation of CO(7-6) (upper-right), a current e-MERLIN observation of CO(1-0) at K-band (lower-right)
      and a second e-MERLIN observation but with the Goonhilly antenna included (lower-left). Each simulated observation
      has 24-hours of on-source time, and colour scales run linearly from the minimum 
      to the maximum pixel value. Contour levels are (-4, -2, 2, 4, 8, 16, 32...) $\times$ 0.01 mJy per beam for the three simulated 
      images and (-4, -2, 2, 4, 8, 16, 32...) $\times$ 0.01 mJy per pixel for the input model. This figure shows that the resolution
      provided by an e-MERLIN array which includes a Goonhilly antenna is sufficient and necessary to resolve and recover
      the disk orientation in a typical z$\sim$4 massive galaxy.
      \label{fig:z4_co}}
\end{figure*}

\subsection{European VLBI (primarily L- and C-band)}

Forming baselines between the existing e-MERLIN telescopes and Goonhilly 
will also improve the overlap between e-MERLIN and the European VLBI 
Network, which has a minimum baseline length of $\sim$270 km (Effelsberg 
- Westerbork\footnote{Note that 12 of the 14 elements which form the 
Westerbork telescope are being fitted with the APERTIF focal-plane array 
system in 2012. Since only two dishes will remain available for VLBI 
work above L-band, the EVN will have greatly reduced C-band sensitivity on this 
baseline.}). Combined EVN+MERLIN observations are currently routine, 
where the shorter baselines of MERLIN are required to resolve larger 
scale structure. Adding a Goonhilly receptor to the e-MERLIN array will 
introduce overlaps in baseline lengths between these two arrays. 
Goonhilly would also increase the western extent of the EVN. For a more 
detailed discussion of this aspect see the article by Kl\"{o}ckner et 
al.~(2011, in prep.).

\subsection{Massive star formation (primarily C- and K-band)}

To understand the complex dynamics driven by gravity, rotation,
radiation and magnetic fields during the process of massive star
formation high resolution probes of both the infalling molecular
material and outflowing ionized gas are needed. ALMA will provide the
required sensitivity to molecular gas at a resolution of around 10--100~mas
which corresponds to 30-300~AU at a typical distance of 3~kpc for
the nearest samples. Only e-MERLIN can provide the matching 
resolution and sensitivity at cm wavelengths to probe the ionized jets
and winds that are an integral part of the star formation process.
	
Significant progress will be made only through joint high resolution
studies of a common sample which necessitates e-MERLIN with Goonhilly
being able to observe equatorial objects also accessible to
ALMA. Furthermore, most massive star formation takes place in the
inner Galaxy (e.g. Urquhart et al.,  2008) and
representative samples require observations in the range -20$^{\circ}$~$<$~Dec~$<$~$+$2$^{\circ}$. Most of the best examples of jets from massive protostars
are found in this region (e.g. Carrasco-Gonzalez et al., 2010; Gibb et al., 2003) and only e-MERLIN with Goonhilly
can provide the factor of ten increase in resolution over current studies to 
probe the inner regions where the flows are being driven and collimated.

\subsection{Resolving molecular gas disks at high redshift (primarily K-band)}

The $^{12}$CO molecule is the best proxy for molecular hydrogen and 
its J(1-0) transition ($\nu_{rest}$~=~115.27~GHz) is redshifted in to 
the K-band of e-MERLIN at z$\sim$4. High-redshift detections of 
molecular gas typically target systems which are undergoing vigourous 
episodes of AGN activity (e.g. quasars) or star-formation (e.g. sub-mm 
galaxies). ALMA will be the natural choice of instrument to target the 
higher-order transitions in such galaxies however it is incapable of 
observing the lowest-order transition which provides the most robust 
estimate of the total mass of molecular gas.

The S$^{3}$-SAX simulation (\cite{obreschkow09}, and references therein) 
makes strong predictions about the cosmological evolution of molecular 
gas disks. We can demonstrate the advantage of adding Goonhilly to this array
by selecting a massive object at z~=~4.023 from the S$^{3}$-SAX database\footnote{{\tt http://s-cubed.physics.ox.ac.uk}}
and simulating a line observation on the celestial equator. Galaxies are extracted from the simulation
catalogue within a random volume of space defined by the field-of-view of a 25-metre
antenna along the spatial axes and the maximum K-band bandwidth along the redshift axis.
A model sky cube is then generated (Levrier et al., 2009) for the most massive galaxy within this volume 
showing the CO(1-0) emission from its molecular disk, redshifted to 22.948 GHz, covered by 32 channels of 21.4 MHz each. For comparison
we also simulate an ALMA observation of the CO(7-6) emission ($\nu_{rest}$~=~806.89~GHz) which is
redshifted into Band 4 of ALMA at 160.63 GHz. A selection of numbered channels from the CO(1-0) cube are shown in the left-hand
column of Figure \ref{fig:z4_co} to illustrate the morphology of the source. 

These two model datacubes are then inverted to generate three sets of visibilities, namely for an e-MERLIN observation both
with and without Goonhilly, and an ALMA observation using a current proposed layout for its most extended configuration. On-source
time in all cases is 24 hours, and noise is added accordingly. The simulations are performed in spectral-line mode and 
imaged with uniform weighting for maximum resolution. The final, deconvolved, frequency-averaged images are presented in Figure \ref{fig:z4_co}. 
Clockwise from the upper-left these are
the input model, the ALMA simulation, the e-MERLIN simulation and the e-MERLIN simulation including Goonhilly. Please refer to the 
caption for contour and colour-scale details.

One of the fundamental requirements for ALMA is excellent image fidelity
and ALMA faithfully recovers the disk structure in our simulation. The current six-element e-MERLIN array barely resolves
the source and because of the beamshape it mis-informs as to the disk alignment. The image produced by e-MERLIN including Goonhilly
comfortably resolves the gas disk. For the three simulated cases this would be the most valuable observation in terms of
constraining the gas kinematics by spatially and spectrally resolving this lowest CO transition in this high-redshift object. 
Note that the resolution of an expanded e-MERLIN array at K-band would exceed that of ALMA for any frequency below Band 7.

Even for galaxies which are barely resolved (e.g.~$\geq$3 independent
points) it is possible to derive a crude rotation curve from the red- and blue-shifted
measurements either side of a mean. Such observations can be combined with models
for gas kinematics in clumpy disks to derive estimates of gas and dark-matter masses
in distant z$\sim$1.5 galaxies (e.g. Daddi et al., 2010).
An expanded e-MERLIN array may well be the best 
facility for extending such studies to z$\sim$4 since obtaining such high spatial
resolution with ALMA would require targeting of CO at impractically high-order
$J$-transitions.

Note also that high-redshift CO is another area in which an extended e-MERLIN would be a unique
pathfinder instrument for influencing the design of the SKA. New and upgraded cm-wave
facilities such as the EVLA (e.g.~Riechers, 2010), Greenbank Telescope (e.g.~Harris, 2010) and eventually MeerKAT (Heywood et al., 2011, these proceedings)
are becoming increasingly important for studies of the low-order transitions of CO at high-z.

A realisation of the SKA with e-MERLIN-like baselines will
be capable not only of detecting and resolving such galaxies, but doing so
with extremely high detection rates via the survey speed offered by the
extreme sensitivity and wider field of view (due to the smaller dish sizes).

\section{Conclusions}

Introducing one of the 30-metre class antennas at the Goonhilly Earth 
Station into e-MERLIN would double the already superb resolving 
power of the array by increasing the maximum baseline from 217~km 
to 441~km.

During the course of an observation an interferometer samples a range of 
spatial scales governed by the baselines between antennas as projected
onto a sky plane perpendicular to the observing direction (thus defining 
the $uv$-coverage), and including Goonhilly results in significant 
improvements in this aspect. Of particular note is how the increased N-S 
extent of the array improves its performance during low-declination 
observations. The point-spread function of an aperture synthesis 
observation is essentially the Fourier transform of the $uv$-coverage, 
and the improved $uv$-coverage when Goonhilly is included results in 
correspondingly lower PSF sidelobes.

An antenna at Goonhilly also increases the overlap in baseline lengths 
between e-MERLIN and the EVN, and would also increase the E-W extent of 
the latter (Kl\"{o}ckner et al., 2011, in prep.).

We have highlighted some science cases where a Goonhilly-enhanced 
e-MERLIN would be an extremely powerful instrument, including resolving 
molecular gas at high-redshift with resolution exceeding that of all but 
the highest ALMA bands, and follow-up observations of newly-discovered 
strong gravitational lens systems. The latter is particularly 
noteworthy as an expanded e-MERLIN may well be the best contemporary 
facility for testing the predictions of the CDM model on small 
mass-scales.

\begin{acknowledgements}
We wish to thank Phil Marshall, Steve Jones and Piran Trezise for useful discussions. We also wish to thank Dave Green
for devising and sharing the ``cube helix'' colour map, an inversion of which is used in Figure \ref{fig:z4_co}.
\end{acknowledgements}


\begin{thebibliography}{}

\bibitem[Blake et al.(2004)]{blake2004} Blake, C.~A., Abdalla, F.~B., Bridle, S.~L., \& Rawlings, S.\ 2004, New Astronomy Reviews, 48, 1063 
\bibitem[Carrasco-Gonz{\'a}lez et al.(2010)]{2010Sci...330.1209C} Carrasco-Gonz{\'a}lez, C., Rodr{\'{\i}}guez, L.~F., Anglada, G., Mart{\'{\i}}, J., Torrelles, J.~M., \& Osorio, M.\ 2010, Science, 330, 1209 
\bibitem[Dalal \& Kochanek (2002)]{dalal} Dalal, N., \& Kochanek, C.~S., 2002, ApJ, 572, 25 
\bibitem[Daddi et al.(2010)]{daddi2010} Daddi, E., et al.\ 2010, \apjl, 714, L118 
\bibitem[Garrett et al.(2010)]{garrett2010} Garrett, M.~A., Cordes, J.~M., Deboer, D.~R., Jonas, J.~L., Rawlings, S., \& Schilizzi, R.~T.\ 2010, arXiv:1008.2871 
\bibitem[Gibb et al.(2003)]{2003MNRAS.339.1011G} Gibb, A.~G., Hoare, M.~G., Little, L.~T., \& Wright, M.~C.~H.\ 2003, \mnras, 339, 1011
\bibitem[Harris et al.(2010)]{harris2010} Harris, A.~I., Baker, A.~J., Zonak, S.~G., Sharon, C.~E., Genzel, R., Rauch, K., Watts, G., \& Creager, R.\ 2010, \apj, 723, 1139 
\bibitem[Heywood et al., (2008)]{heywood} Heywood, I., Kl\"{o}ckner, H.-R., \& Rawlings, S.\ 2008, arXiv:0801.2037 
\bibitem[Heywood et al., (2011)]{heywdoo11} Heywood I., et al., 2011, arXiv:1103.0862
\bibitem[Levrier et al., (2009)]{2009arXiv0911.4611L} Levrier, F., Wilman, R.~J., Obreschkow, D., Kloeckner, H.~-., Heywood, I., \& Rawlings, S.\ 2009, arXiv:0911.4611 
\bibitem[Negrello et al., 2010]{negrello} Negrello, M.~et al., 2010, Science, 330, 800
\bibitem[Obreschkow et al., (2009)]{obreschkow09} Obreschkow, D., Kl{\"o}ckner, H.-R., Heywood, I., Levrier, F., \& Rawlings, S.\ 2009, \apj, 703, 1890 
\bibitem[Patel et al.(2010)]{2010MNRAS.401.2572P} Patel, P., Bacon, D.~J., Beswick, R.~J., Muxlow, T.~W.~B., \& Hoyle, B.\ 2010, \mnras, 401, 2572 
\bibitem[Riechers et al.(2010)]{riechers2010} Riechers, D.~A., Carilli, C.~L., Walter, F., \& Momjian, E.\ 2010, \apjl, 724, L153 
\bibitem[Urquhart et al.(2008)]{2008A&A...487..253U} Urquhart, J.~S., et al.\ 2008, \aap, 487, 253 
\bibitem[Vegetti \& Koopmans (2009)]{vegetti} Vegetti, S. \& Koopmans, L.~V.~E., 2009, MNRAS, 400, 1583

\end{thebibliography}
\end{document}